\documentstyle[12pt]{article}

\newcommand\beq{\begin{equation}}
\newcommand\eeq{\end{equation}}
\def\beqa{\begin{eqnarray}}
\def\eeqa{\end{eqnarray}}
\def\bega{\begin{array}}
\def\enda{\end{array}}

\def\non{\nonumber}

\def\q{{\bf q}}
\def\p{{\bf p}}
\def\bq{Q}
\def\bp{P}
\def\H{{\cal H}}
\def\bH{{\overline{\cal H}}}
\def\la{{\lambda}}
\def\LLa{{(\Lambda-\Lambda_0)}}
\def\La{{\Lambda-\Lambda_0}}
\def\t{{t}}

\begin{document}
\title{Evolving Constants of Motion}

\author{Arlen Anderson\thanks{arley@physics.unc.edu}\\
Dept. Physics\\ UNC-Chapel Hill\\  Chapel Hill NC 27599-3255.}
\date{July 17, 1995}

\maketitle
\vspace{-8cm}
\hfill IFP-UNC-95-511

\hfill gr-qc/9507038
\vspace{7cm}

\begin{abstract}
A critical presentation of Rovelli's ``evolving constants of motion''
is given.  Previous criticisms by Kucha\v{r} concerning the role of factor
ordering and the non-existence of observables are dealt with and
shown to be unfounded.  Kucha\v{r}'s criticisms that this approach
does not solve the global, multiple choice or Hilbert space problems
of time are confirmed, and new insight into why this is so is
obtained.
\end{abstract}
\newpage

Rovelli\cite{Rov1} has proposed that one can interpret ``timeless''
parametrized theories, such as canonical quantum gravity, in terms of
``evolving constants of motion.'' Kucha\v{r} has criticized this approach on a
number of grounds, from issues of factor ordering\cite{Kuc1} to the
existence of observables\cite{Kuc2}, but, as a means of working with
parametrized theories, there is much to recommend it.
Some important conclusions can be learned from an independent critical
analysis.

In this paper, Rovelli's approach will be shown to be essentially the
Heisenberg picture as formulated in a parametrized theory. The evolving
constants of motion are observables which commute with the
super-Hamiltonian\cite{And1}, and they can be constructed from solutions of
the Heisenberg equations of motion. One does not have to solve a factor
ordering problem to find them. Existence of such solutions and hence of
observables is guaranteed, and Kucha\v{r}'s doubts\cite{Kuc2} to the contrary
are unfounded. On the other hand, Rovelli's program does fail to solve many
of the problems of time\cite{Kuc1}, including the global problem of time,
the multiple choice problem and the Hilbert space problem. The reason is
that while the solutions to the Heisenberg equations of motion can be found
without specifying the Hilbert space structure of a theory and without
specifying a foliation of spacetime, to have a complete quantum theory,
these must be specified and the observables do not determine them.
In particular, self-adjointness of observables is
insufficient to determine the inner product.

\section{Rovelli's evolving constants of motion}

Rovelli's approach\cite{Rov1} is perhaps best described in the context of
parametrized quantum mechanics. This allows the freedom to be specific in
examples without worrying about the particular structure of something like
the Wheeler-DeWitt equation. One begins by working in an extended phase
space which contains time and its conjugate momentum as well as the spatial
phase space variables. I use bold face letters $(\q,\p)$ to denote the
collection of extended phase space variables, $(q_0, p_0)$ for time and its
conjugate momentum, and $(q,p)$ to denote the spatial phase space
variables. It is convenient to work as if there were only one spatial
degree of freedom, but this is notational. I assume that the extended phase
space has the topology $R^{2n+2}$, where $n$ is the number of spatial
degrees of freedom, and that the extended phase space variables satisfy the
canonical commutation relations: in an obvious index notation,
\beq
[q_\mu,p_\nu]=i\delta_{\mu\nu},\quad [q_\mu,q_\nu]=0,\quad [p_\mu,p_\nu]=0.
\eeq

Consider a classical action in parametrized form
\beq
S=\int d\lambda \,\biggl(\p \dot \q-N(\la)\H_{\rm cl}(\q,\p)\biggr),
\eeq
where $\H_{\rm cl}(\q,\p)$ is called the super-Hamiltonian, $N(\lambda)$
will be called the lapse, and the dot indicates differentiation with
respect to the affine parameter $\la$. In non-relativistic physics, the
super-Hamiltonian is
\beq
\H_{\rm cl}(\q,\p)=p_0+H_{\rm cl}(q,p,q_0).
\eeq
The ``super'' prefix distinguishes it from the familiar Hamiltonian
$H_{\rm cl}(q,p,q_0)$.

The equations of motion are found by variation.  Variation of the
lapse gives the super-Hamiltonian constraint
\beq
\H_{\rm cl}(\q,\p)=0.
\eeq
This constraint is the signature of a parametrized theory.  Varying the
extended phase space variables gives the Hamilton equations of motion
\beq
\label{cleom}
\dot q_\mu=N(\la)\{q_\mu,\H_{\rm cl}\},\quad
\dot p_\mu=N(\la)\{p_\mu,\H_{\rm cl}\}.
\eeq

When this system is quantized, one finds the (operator-valued)
super-Hamiltonian constraint
\beq
\H(\q,\p)=0.
\eeq
and the Heisenberg equations of motion
\beq
\label{Heom}
\dot q_\mu=-iN(\la)[q_\mu,\H],\quad \dot p_\mu=-iN(\la)[p_\mu,\H].
\eeq
Kucha\v{r}'s first objection\cite{Kuc1} is that there is more than one way to
quantize a given classical system to obtain the quantum $\H(\q,\p)$.
This is certainly true, but it is the generic problem one faces when
quantizing anything.  At the present time, there is no answer to
this ``multiple choice'' problem, except by appeal to experiment.  As such,
it is a question about choosing between candidate quantum
theories and not an obstruction to formulating them.  I assume that a
factor ordering of the super-Hamiltonian has been given.

The presence of the super-Hamiltonian constraint raises the so-called
problem of the frozen formalism\cite{fro}. In a theory with (first-class)
constraints, observables are defined to be functions which commute (weakly)
with all of the constraints. This is necessary because otherwise the action
of an observable would take one out of the constraint hypersurface. But
here, the super-Hamiltonian would seem to be the generator of time
translations, so that if an observable commutes with $\H$, it must be a
constant of the motion. Where then are the dynamics one expects of
observables? Rovelli\cite{Rov1} introduced the ``evolving constants of
motion'' to resolve the apparent paradox. The multiple conflicting uses of
the word ``observable'' were discussed in \cite{And1}, and a simple
example was done to show that, properly understood, there is in fact no
problem.

The resolution is based on the recognition that
observables are members of families of constants of motion parametrized by
a label related to time.  Given a self-adjoint operator to be measured, it
generically has a realization as a distinct observable at each instant of
time. Its dynamical evolution is then a reflection of its motion through
the family of constants of motion with the passage of time.

To be concrete, consider the elementary example of
the parametrized free particle
\beq
\H=p_0+{1\over 2} p^2
\eeq
The collection of operators,
\beq
\label{Q}
Q(\t)=q+p(\t-q_0),
\eeq
parametrized by $\t$, are easily verified to be observables
\beq
[Q(\t),\H]=0.
\eeq

To interpret these observables correctly, one must pass through an
unfamiliar step.  In the parametrized formalism, $q_0$ is an operator.
When it acts on a state defined on a spacelike slice at a fixed instant
of time, its eigenvalue is the time associated to the slice on which that
state is defined, thus
\beq
\label{q0ev}
q_0|\psi_1,t_1\rangle=t_1|\psi_1,t_1\rangle.
\eeq
The Hilbert space structure of the theory is still defined at fixed moments
of time.  One cannot form superpositions of states at different moments
of time---the instants of time define superselection sectors.
This means that $q_0$ is not integrated over in the Hilbert space inner
product.

While $q_0$ is a self-adjoint operator (its eigenvalues are
all real),  it is not an ``observable'' in the practical sense of ``something
one observes/measures.''  One cannot build devices which couple to
it directly.  Thus, while a coupling like $H_I=c q_0 q$
is acceptable mathematically, physically we can only build devices which
couple the (by definition, spatial) degrees of freedom of different
subsystems.  A
coupling like $H_I$ may arise indirectly as a consequence of the detailed
evolution of some physical degree of freedom, but the true coupling is
between spatial degrees of freedom.  Finally, and perhaps most
importantly, $q_0$ does not commutes with
the super-Hamiltonian.  It is not an observable.

When the observable $Q(\t)$ acts on a state, one obtains
\beq
\langle \psi_1,t_1| Q(\t)|\psi_1,t_1\rangle=\langle
\psi_1,t_1|q+p(\t-t_1)|\psi_1,t_1\rangle.
\eeq
Thus, $Q(t_1)=q$ when acting on states defined on the slice at time $t_1$.
It is essential to emphasize that one must give the time of the state that the
observable acts on to know its behavior.  The operator (\ref{Q}) is not
physically meaningful on its own; it acquires meaning in conjunction with
the states it acts on.  Another way to say this is that an operator does
not have physical meaning until one specifies the Hilbert space in which
it acts.

Suppose one wishes to compute
the expectation value of $q$ at time $t_2$ in terms of states at time
$t_1$.  One pulls back the operator $q$ from time $t_2$ to time $t_1$
and finds that one has the Heisenberg evolution, based on states at
time $t_1$:
\beqa
\langle \psi_2,t_2|q|\psi_2,t_2\rangle&=&
\langle \psi_1,t_1|e^{ip^2(t_2-t_1)/2}q e^{-ip^2(t_2-t_1)/2}
|\psi_1,t_1\rangle \\
&=&
\langle \psi_1,t_1| q+p(t_2-t_1)|\psi_1,t_1\rangle. \nonumber
\eeqa
But, this is just
\beq
\langle \psi_2,t_2|Q(t_2)|\psi_2,t_2\rangle=\langle \psi_1,t_1|
Q(t_2)|\psi_1,t_1\rangle.
\eeq
This confirms what we have been told:  $Q(t_2)$ is a constant of the
motion; it has the same expectation value on every slice.

Here, we have held the observable fixed and varied the slice.  Turning
the story around, if we hold the slice and state fixed, say at $t_1$,
and vary the observable, then, as a function of $t$, we have
\beqa
\langle \psi_1,t_1|Q(t)|\psi_1,t_1\rangle&=&\langle \psi_1,t_1|
q+p(t-t_1)|\psi_1,t_1\rangle\\
&=& \langle \psi_1,t_1|e^{ip^2(t-t_1)/2}q e^{-ip^2(t-t_1)/2}
|\psi_1,t_1\rangle .  \nonumber
\eeqa
Thus, $Q(t)$ gives the Heisenberg evolution of $q$ from $t_1$ to $t$
when acting on states at $t_1$! (More generally, $Q(t)$ acting on states at
time $\tau$ is the Heisenberg
evolution of $q$  from $\tau$ to $t$.)
Dynamical evolution is the movement of the self-adjoint operator under
observation through the family of constants of motion as time passes.

To be complete, suppose one is in the Heisenberg picture with states
defined at time $t_1$,  and one wants to follow the Heisenberg evolution of
the operator $q+p(t_2-t_1)$, corresponding to $Q(t_2)$ acting on
states at time $t_1$.   One can decompose $Q(t_2)$ in terms of observables
at time $t_1$ by
\beqa
Q(t_2)&=&e^{iP^2 (t_2-t_1)/2} Q(t_1) e^{-iP^2 (t_2-t_1)/2} \\
&=&Q(t_1)+(t_2-t_1)P. \nonumber
\eeqa
(where $P=p$ for all $t$).
The Heisenberg evolution of $q+p(t_2-t_1)$ is then
\beq
\langle \psi_1,t_1|Q(t)+(t_2-t_1)P|\psi_1,t_1\rangle.
\eeq

In the general case of non-relativistic quantum mechanics with
time-independent Hamiltonian,  the super-Hamiltonian is
\beq
\H=p_0+H(q,p).
\eeq
Any self-adjoint operator $f=f(q,p)$ can be promoted to a
family of observables by computing its Heisenberg evolution
\beq
F(\t;q,p,q_0)=e^{iH(\t-q_0)}f(q,p)e^{-iH(\t-q_0)}.
\eeq
It is easily verified that
\beq
[F(\t;q,p,q_0),\H]=0.
\eeq
The expectation value of the operator $f(q,p)$ at time $t_1$ corresponds to
the expectation value of the observable $F(t_1;q,p,q_0)$. Viewed as a function
of $t$, $F(t;q,p,q_0)$ acting on states at time $t_1$ is simply the
Heisenberg evolution of $f(q,p)$ from time $t_1$ to $t$. The
novel feature of the observable $F(t;q,p,q_0)$ is that its dependence on
the operator $q_0$ allows one to shift the fixed initial slice on which the
Heisenberg picture is defined. In a sense, some of the freedom of the
Schrodinger picture to change slices is incorporated, though dynamics still
remains with evolution of the operator. Here, that evolution is seen to be
movement through the family of observables as time passes.

Note that $q_0$ cannot be promoted to an observable in this way.  From
(\ref{Q}), one can construct an operator
\beq
\bq_0={1\over p}(\bq-q) +q_0
\eeq
which commutes with ${\cal H}$---but it is not self-adjoint in the usual
inner product.

Consider two of Kucha\v{r}'s remarks\cite{Kuc2}:  a) ``one can observe
dynamical variables which are not perennial;'' b) ``perennials are
often difficult to observe.''  (``Perennial'' is synonymous with
observable as used here.)  Neither of these remarks is truly in conflict
with what has just been described.  On first consideration, the self-adjoint
operator $q$ (at time $t_1$) is measurable\cite{meas1}, and it does not commute
with the super-Hamiltonian, so it is not an observable/perennial.  Kucha\v{r}'s
statement is apparently correct.  Rovelli takes one small further step.  He
recognizes that at time $t_1$, $q=Q(t_1)$, that is, it is an instance
of an observable.  In Rovelli's scheme, every dynamical variable at an
instant is simply the instantaneous form of some observable.  This seems
a modest step, but it saves one the mental
gymnastics of coping with operators which may take one out of the constraint
surface, by assuring that they never do.

Kucha\v{r}'s second remark is a fact.  If one attempts to observe $Q(t_2)$
at time $t_1$, this may be difficult.  Fortunately, we don't
experience time in the Heisenberg picture, and there is nothing which
says we must try to do so.  At each instant of time, there are a collection
of observables which are comparatively easy to measure, and these are
the ones our attention focuses on.  Other observables may be difficult
to measure, and our effort to determine them will depend on our interest
in them.

This brings us to Kucha\v{r}'s key question.  He asks, how is $\t$
to be observed?  In the present context, this
question is somewhat misdirected, but it is a very important question
in its place
and is discussed at length in \cite{Andother}.
Ostensibly, Kucha\v{r} wants to argue that to
observe change, one must measure something which is not an observable.
He argues that one must know $\t$ to know when to measure a particular
observable.
On the face of it, this seems reasonable, but it puts a false emphasis on
$\t$ and in doing so misses an essential point.

One does not choose to measure $Q(\t)$ at some instant because one knows
that instant is labelled by $\t$.  No, one chooses to measure an
operator, say $q$, at the instant which is ``now,'' and there is an
observable associated to this operator---which observable, in an absolute
sense, being irrelevant.  At a different instant, one again measures $q$,
now associated to a different observable, and generally the values
obtained are different.  Change has been measured.  It is true that there
is a hidden active agent in the passage of time, but it is enough that it
happens.  It is inferred only indirectly by the fact that the value of the
observable changes.

Alternatively, as a theorist outside of the system, one can
select a slice and measure $q$ there.  It gives a value associated to
one observable.  When $q$ is measured again on a different slice,
a distinct value associated to another observable is obtained.  Change
has been measured.

Generally, we want more than to show that observables take
different values.  We want to coordinate the change in those values.
It is not sufficient that time pass; we must mark the passage of time.
This cannot be done in the model at hand.  It has only  one
simultaneously measurable observable.
Kucha\v{r} asserts that $q_0$ is the hand of an ideal Newtonian clock.  This is
false: $q_0$ is not measurable because it is not a degree of freedom of a
physical subsystem.

The traditional approach is to introduce an additional degree of freedom,
say with Hamiltonian $H=p_x$, and call it a clock.  This Hamiltonian is
used here solely to illustrate a point.
As it is not bounded from below, it is
unphysical, but consider it temporarily to be a measurable
subsystem.  (Clocks are critically discussed more
fully in \cite{Andother}.) A second family
of observables (simultaneously measurable with $Q(\t)$) can be constructed
\beq
X(\t)=x+\t-q_0.
\eeq
$X(\t)$ is
the hand of the clock, and it can be used to coordinate different
measurements of $Q(\t)$.

Suppose that at some initial instant, arbitrarily labelled $t_1$, the
$x$-sub\-system is in an eigenstate of $X(t_1)$ with eigenvalue $x_1$. The
state of the full system at $t_1$ is then $|\psi_1(q),x_1,t_1\rangle$. The
state of the $x$-subsystem at a later time $t_2$ can be deduced from the
constancy of the observable $X(t_2)$. Computed at time $t_1$, one finds
\beqa
\langle \psi_1, x_1,t_1|X(t_2)|\psi_1,x_1,t_1\rangle &=& x_1+t_2-t_1 \\
&=& \langle \psi_2, x_1+t_2-t_1,t_2|X(t_2)|\psi_1,x_1+t_2-t_1,t_2\rangle.
\nonumber
\eeqa
This is what one expects from Heisenberg evolution
from $t_1$ to $t_2$ using the $H=p_x$.

By repeatedly measuring $x$ as time passes, one can wait a predetermined
$x$-interval between measurements of $Q(\t)$. One need never know the
dependence of $Q(\t)$ on $\t$ because one will be able to infer a
dependence on $X(\t)$ instead. It should be repeated that $X(\t)$ refers to
the collection of outcomes of measurements of $x$---the label $\t$ is
largely a convenience to distinguish between different values of the
outcomes. Again, behind the whole process, it is assumed that ``time
passes.'' This may be justified either by appeal to experience or by saying
that this is the phrase which describes the process of sequentially
considering the states associated with different values of $X(\t)$. While
$X(\t)$ can be measured on any slice, it is natural to measure $x$ on some
slice and assign $X(\t)$ to the value obtained there. Having done so, one
may measure $q$ on the same slice and assign $Q(\t)$ to that value.

The point here is that the passage of time is an external experiential, if
you will, phenomenon. The evolving constants of motion do not explain the
passage of time. They show that when one observes measurable quantities
like $q$ at a sequence of experienced instants of time $t_k$, these
observations can be
understood as the measurement of observables $Q(t_k)$ which commute with
the super-Hamiltonian ${\cal H}$. To some, the use of observables which
commute with ${\cal H}$ may seem unnecessary, and they will say one is
really measuring $q$ which does not commute with ${\cal H}$. It is not that
one cannot do physics in this way. It is a question of whether it is in
keeping with the spirit of other tenets of mathematics and quantum physics
where one is ill-disposed to operators which can take one out of the
Hilbert space or constraint surface.

\section{Observables and Integrability}

Another of Kucha\v{r}'s arguments against Rovelli's evolving constants of
motion
concerns the existence of observables that commute
with the super-Hamiltonian constraint. Kucha\v{r}\cite{Kuc2} reminds
us of Poincar\'e's
results on the nonexistence of further integrals of the motion beyond the
classical integrals (energy, center of mass momentum, and angular momentum)
for systems like the asymmetric top and the three-body problem. He goes on
to argue that gravity itself is unlikely to be integrable (cf.
\cite{Tor}) and therefore there will not be a complete set of observables
with which to characterize a system. There is a confusion of terms in this
objection, with Kucha\v{r} and Rovelli talking past each other.

The problem is with what it means to be a constant of motion. Both Kucha\v{r}
and Rovelli agree at the outset that an observable which commutes with the
super-Hamiltonian is a constant of motion. The difficulty follows as
Kucha\v{r}\cite{Kuc2}
begins to speak of what are often referred to as first integrals of the
motion while Rovelli speaks of what may be called ``second
integrals'' of the motion\cite{Coe}, but are more readily recognized as
solutions of the equation of motion. When integrating a system of
second order differential equations, such as equations of motion,
one integration may reduce the order
of some equation by one and one obtains a ``first integral.'' The energy
and the
momenta of ignorable coordinates are obtained by
such a first integration of the Euler-Lagrange equations. These are all
first integrals of the motion. A second integration produces a
``second integral,'' i.e. the solution of the equation.

Since there are existence and uniqueness theorems for the solutions of the
equations of motion of dynamical systems, including Einstein's field
equations, the existence of second integrals is not in doubt. Each solution
is given uniquely in terms of its initial data, and it is considered a
trivial observation that these initial data are constants of motion for
that solution. There is a canonical transformation which takes one from any
point along a solution back to its initial conditions, so there are
canonical transformations which trivialize the motion of the system. The
problem with second integrals of the motion is finding them! That is the
central challenge of dynamics: solving the equations of motion.
This is why first integrals of motion are so valuable. Their existence
assists in the integration of the equation of motion.

An alternative definition of a first integral is as a function of the
positions and velocities which is constant, $A({q}(t),{\dot q}(t))={\rm
constant}$, along every solution of the equations of motion. It appears that
the key feature that distinguishes such an integral from a second integral
is that the function $A$ has no explicit dependence on time, but this is
misleading. Kucha\v{r} attempts to confuse the distinction between first and
second integrals by fixing the energy. The solutions to the equations of
motion at fixed energy do not involve the time, so it seems by this
definition any integrals must be first integrals. This is of course absurd.
A non-integrable system with time-independent Hamiltonian for which one has
the exact time-dependent solution for a given set of initial conditions
does not acquire first integrals simply because one fixes the energy nor
does the solution cease to exist.

The fixed-energy second integrals are
found by inverting the solution of one of the variables for the time
and using this to eliminate the time in the remaining solutions of the
other variables.  It is clear that, when one does this, the expressions
for the other variables will involve both initial and final values of
the variable which has been used to eliminate time.  A true first
integral however only depends on initial variables.  In an example in the
next section,  (\ref{obs1}) is a first
integral while (\ref{obs3}) is only a second integral.
This is the essential distinction between
first and second integrals.  In the usual context, it is the time which
occurs as a difference of final and initial values, and that is why second
integrals are associated with time-dependence.

A system of $n$ degrees of freedom is said to be integrable if it admits
$n$ first integrals of the motion in involution.  Such a system is
equivalent under a time-independent canonical transformation to a system
constrained to move geodesically on a regular (flat) torus.  It should be
emphasized that to say a system is non-integrable does not mean that the
solutions to its equations of motion cannot be characterized.  It means
that they are not equivalent to geodesic motion on a torus.  Thus,
geodesic motion on a two-dimensional higher genus Riemann surface of
constant negative
curvature is easily described, for example by circular arcs in the
Poincar\'e disk tesselated by the fundamental domain appropriate to the
surface, but the motion is non-integrable.

The evidence is that general relativity is not an integrable theory, and
that there are few if any first integrals\cite{Tor}. This is not really
surprising. It is in fact a very good thing. I interpret it in the
following way. Since general relativity is not integrable, there is not a
time-independent canonical transformation to variables in which all the
momenta are constant and their conjugate variables evolve linearly in time.
This means that the universe cannot be in a superposition of eigenstates of
a complete collection of $t$-independent observables
(where $t$ is a final value of the time). As a
result, time exists. We cannot be trapped in the frozen formalism.

The foundation of Rovelli's approach is the observation that nevertheless
the universe is in a superposition of eigenstates of a complete collection
of $t$-dependent observables. Each of these observables are simply the
Heisenberg evolution of some functions of the initial data. If we choose
to work in the Heisenberg picture, the universe is frozen in a particular
state. At each instant of time, however, we must use a different collection
of observables to decompose the state of the universe. This is because we
must use a collection of observables evaluated at that moment of time and
therefore necessarily a collection of time-independent observables. Since
general relativity is not integrable, we cannot use the same set of
time-independent observables at every time, so our set must change. In the
Heisenberg picture, it is the changing of the set of observables that we
use to decompose the state of the universe which reflects the passage of
time.

\section{The Problems of Time}

Returning to the general situation, the challenge in constructing Rovelli's
evolving constants of motion is to solve the Heisenberg equations of
motion. Kucha\v{r}\cite{Kuc1} raises several objections about one's ability to
do this, but the immediate form of his objections rest largely on a
misapprehension of the task
one faces. Kucha\v{r} approaches the problem of construction as one of factor
ordering a classical solution of the problem. After emphasizing the
difficulty of factor ordering
in general, he goes on to ridicule the existence of a time function
which he argues is a necessary part of the procedure. He
points out the ``global problem of time''---that generally there are
obstructions to foliating extended phase space with a single time
function---and the ``multiple choice problem''---that one could use
different time functions and one must prove that the quantum theories are
equivalent. He argues that Rovelli's hope to solve the Hilbert space
problem, that is to put a Hilbert space structure on the space of solutions
to the super-Hamiltonian constraint, is in vain because of the multiple
choice problem. He concludes that Rovelli cannot solve any of the problems
of time.

The situation is not quite so grim as this, though the conclusions are
valid. Kucha\v{r}'s version of
Rovelli's approach is a straw man which distorts both the strengths and
weaknesses of a careful treatment. The role of the time function
is overemphasized, and
one especially does not have to factor order a classical solution. One
simply needs to solve the Heisenberg equations of motion (\ref{Heom}).
Doing so may be non-trivial, but existence and uniqueness of the solution
are assured. Perturbative expansions in $\La=\int^\la_{\la_0} N(\la) d\la$
of the
solution are easily obtained in the finite dimensional case, and quantum
canonical transformation methods are under development to obtain
expressions with well-defined factor ordering in more closed
form\cite{And2}.

van Hove's theorem may haunt the passage between classical and quantum
theory, but it is impotent if one does not bother with the crossing and
works always in the quantum realm. The real difficulties are serious enough
without adding to them. For example, it is all too true that the field
theory case is of a different magnitude of difficulty than finite
dimensional examples. The functional nature of $N(\la)$ and the consequent
notion of ``bubble-time'' necessarily complicates the production and
representation of solutions. It is not an overstatement to say that
Rovelli's approach is in a state of development, but a less polemical
critique may clarify the issues and lead to further progress.

To obtain reparameterization-invariant observables, one must eliminate the
affine parameter $\La$ in the
expressions for the solutions, and this raises the first serious obstacle.
(A different procedure involving an averaging over $\La$ is discussed
in \cite{Mar1} and may avoid this difficulty.)
To isolate $\La$, one must invert the solution of one variable for $\La$.
One may of course perform canonical transformations on the variables
before attempting the inversion.  There is unfortunately no non-abelian
form of the implicit function theorem of which I'm aware, so it is not
clear when this can be done, though some theorem seems likely to be true.
Additionally, inversion is generally not unique, because of branch choices
or other non-analytic behavior. This means that not every solution of the
equations of motion has the same form in terms of the deparametrized
variables.

As a simple example, consider the super-Hamiltonian
\beq
\H=-{1\over 2}p_0^2-a q_0+ {1\over 2} p^2 + b q.
\eeq
Let $\bq_\mu$, $\bp_\mu$ denote the evolved
variables and $q_\mu=q_\mu(\Lambda_0)$, $p_\mu=p_\mu(\Lambda_0)$ denote
the initial
conditions.  The Heisenberg equations of motion are then
\beq
\dot \bq_\mu=-i[\bq_\mu,\bH],\quad \dot \bp_\mu=-i[\bp_\mu,\bH],
\eeq
where $\bH$ is the super-Hamiltonian expressed in terms of the evolved
variables $\bq_\mu,\ \bp_\mu$, and dot means differentiation with respect
to $\Lambda$.  The solution of these equations are easily
found
\beqa
\label{sol1}
\bp_0 &=& a \LLa +p_0, \\
\bq_0 &=&-{1\over 2}a (\La)^2 - p_0 \LLa +q_0, \non \\
\bp &=& -b \LLa +p, \non \\
\bq &=&-{1\over 2}b (\La)^2 + p \LLa +q. \non
\eeqa

Solving for $\La$ in each of these, one finds the expressions
\beqa
\label{La1}
\La&=&{\bp_0-p_0\over a} \\
&=& -{p_0\over a} \pm {1\over a} (p_0^2 + 2a(q_0-\bq_0))^{1/2}  \non \\
&=& -{\bp-p\over b}  \non \\
&=& {p\over b} \pm {1\over b} (p^2 + 2b(q-\bq))^{1/2} \non
\eeqa
The quadratic equations are solved by shifting $\La$ to cancel the term
linear in $\La$ and then taking the square root of the remaining
expression.  When solving for $\La$, it is assumed to commute with
all of the variables, and
there are no ordering problems with its coefficients.  On the other hand,
the evolved variables do not commute with the initial ones;
the various commutators can be
deduced by taking commutators with the solutions (\ref{sol1}).  In this
example, the forms of $\La$ are the same as they would be classically
(but with appropriate operator ordering of the square-roots).

Equating the first two of (\ref{La1}), one has
\beq
\label{bp0}
\bp_0=\pm \biggl(p_0^2 + 2a(q_0-\bq_0) \biggr)^{1/2}.
\eeq
Classically it is clear why there are two solutions:  for appropriate
initial conditions, a particle climbing
in a linear potential passes a given point once on its way up and again on the
way down.  If one uses the second form of (\ref{La1}) to eliminate $\La$
in either $\bp$ or $\bq$, one will also get two solutions.  Only if one
uses either the first or third forms of (\ref{La1}) will one find
unique forms for the deparametrized solutions, and that is because these
forms are linearly correlated to $\La$.

Observables are formed by substituting the form of $\La$ obtained from one
solution in the solutions
for the other variables.  Here, one treats the evolved and initial
variables as commuting.  Each expression consisting of a
function of evolved variables equated to a function of initial and evolved
variables is then an observable.  For example, use the second form of
$\La$ from (\ref{La1}). Substituting this in the first solution of
(\ref{sol1}), one has (\ref{bp0}) as an observable or alternatively
\beq
\label{obs1}
{1\over 2}\bp_0^2+a\bq_0={1\over 2} p_0^2+a q_0
\eeq
which one can verify satisfies
\beq
\protect[{1\over 2}\bp_0^2+a\bq_0,\H]=0.
\eeq
Note that this is a first integral because the initial and final
variables are separated.

Substituting the second form of $\La$ from (\ref{La1}) in the third
solution of (\ref{sol1}), one finds
\beq
\label{obs3}
\bp= {b p_0\over a} \pm {b\over a} \biggl( p_0^2 +2a(q_0-\bq_0)
\biggr)^{1/2} +p,
\eeq
which can be confirmed to satisfy $[\bp,\H]=0$.  This is a second
integral because $\bq_0$ is present on the right-hand side and cannot be
separated from the initial variables. As well, the more complicated
expression obtained for $\bq$ also satisfies $[\bq,\H]=0$.  Note that
because  $\bp$ and $\bq$ depend on $\bq_0$, they are actually
families of observables parametrized by the value of $\bq_0$.
($\bq_0$ fulfills the role that $t$ played above.)
Further
observables can be found by eliminating other forms of $\La$ and by taking
functional combinations of existing observables.  Obviously,
not all observables commute nor are they necessarily independent.

The multivaluedness of the expressions for $\La$ is a reflection of the
global problem of time.  When none of the variables  has a single-valued
expression in terms of $\La$, then there is an obstruction to
defining a global time function (cf. \cite{And3} for one possible way to
deal with this). There is an important issue here which deserves closer
study, but it is not
altogether clear that it is a problem.  For example, in the present
example, $p_0$ is linearly correlated to $\La$ and would therefore be a good
time function.  Physics however might dictate that $q_0$ is the time.
One has then some sort of time-dependent relativistic super-Hamiltonian and
it is not clear why there is a problem if the classical or Heisenberg
solutions to the equations of
motion are not single-valued in $q_0$.  Further work is needed on this
point.

Having found a complete set of observables, Rovelli's next step is to
determine the inner product by choosing one in which all of the observables
are self-adjoint. If he were successful in this, he would solve the Hilbert
space problem; he would succeed in putting a Hilbert space structure on the
solutions of the super--Hamiltonian constraint. Kucha\v{r} claims he must fail
because of a variant on the multiple choice problem: he can choose
different time functions to construct his observables, and for each he gets
a different set of observables; if one requires that all of these
observables be self-adjoint in the same inner product, there are bound to
be problems. In the example above, different ways to eliminate $\La$
correspond to different choices of Kucha\v{r}'s time function.
While related to the real issue, this argument may lead one to
miss the crucial point.

There is indeed a multiple choice problem which Rovelli has
overlooked which defeats this part of his program.  So far we have the
solutions to the Heisenberg equations of motion, but these solutions are
independent of the Hilbert space structure.  It has not been emphasized,
but the procedure for solving the Heisenberg equations of motion
either perturbatively or by canonical transformation\cite{And2} involves
the algebra of the commutation relations but not the inner product or the
states.  In the solutions, the
spatial and temporal variables are on an equal footing and are
indistinguishable.  The symmetry between spatial and temporal variables
is broken however when one chooses how to make the time-slice through extended
configuration space on which one normalizes states.  There is nothing
except association with physical quantities and their experimental behavior
to guide this choice.  Simple inspection of
the mathematical variables is insufficient.  There is a different inner
product in which all of the observables are self-adjoint for nearly every
choice of slice through the extended configuration space.

As a simple example illustrative of this ambiguity, consider the
massless relativistic free particle,
\beq
\label{free}
\H=-p_0^2+p^2=0.
\eeq
The solution to the parametrized Heisenberg equations of motion are
\beq
\bq=2p\LLa+q,\quad
\bq_0=-2p_0 \LLa+q_0,
\eeq
with $P=p$ and $P_0=p_0$ constant.
There is clearly symmetry between $q$ and $q_0$ in these equations.
Eliminating $\La$ by solving for it in terms of $q_0$ gives
\beq
\label{qeq}
\bq(\bq_0)=-{p\over p_0}(\bq_0-q_0)+q.
\eeq
This system has one physical degree of freedom and requires two
observables.  They can be chosen to be $P$ and $\bq$.
Requiring that they both be self-adjoint on a hypersurface
of constant $t$ [the eigenvalue of $q_0$ as in (\ref{q0ev})]
in a Hermitian inner product, one would find an inner product.
On the other hand, one could solve (\ref{qeq}) for $\bq_0(\bq)$, and
with $P_0$, one could find a different inner product integrated over
a hypersurface of constant $q$.  The choice of hypersurface distinguishes
these from the infinitely many other possibilities, and the correct
choice is dictated by physics.  The requirement
that the observable be self-adjoint is not enough.

In the discussion in \cite{Ash} of the analogous case of a massive
relativistic free particle, it is argued that a unique inner product is
found when one requires the boost operator $\bq_0 P -\bq P_0= q_0 p -q
p_0$ (and say $P$) be self-adjoint.  [This is just another rewriting of
the observable (\ref{qeq}).]  A hidden choice has been made when the inner
product takes the form of an integral over a surface at constant $p_0$.
In the general case of the Wheeler-DeWitt equation, one does
not know which variable in the super-Hamiltonian is time, and one does
not know how to select the hypersurface on which to define the inner product.
The example here makes this obvious.  The only {\it a
priori} justification one has to consider an inner product integrated
over slices of constant $q_0$ is the historical convention that the
variable with the subscript $0$ is time.  I could however have maliciously
mislabelled the variables, and $q$ might be the physical time.
(For completeness, I recall also that
it is well known that
one cannot use the signature of the operator to help to identify time
because the signature of the Wheeler-DeWitt operator is
unrelated to time.)
 Just as
in the classical theory, inspection of the solutions of the equations of
motion is insufficient to determine which variable is time.

Incidentally, the global problem of time seems to have yet another
opportunity to make an appearance.  Having chosen one slice on which to
define one's inner product, it is not clear that this slice can be
extended into a foliation.  It is difficult to say more without specific
examples, but the possibility would seem to exist.

It may have been too optimistic to hope that the Hilbert space problem would
be solved by the evolving constants of the motion, but this should not
detract from the other real accomplishments of this approach.  The
Heisenberg picture is brought to the fore, and the solutions of the
Heisenberg equations of motion are seen not to require a time function
for their definition.  They provide a system of observables which commute
with the super-Hamiltonian, and through their behavior dynamics can be
analyzed.  Space and time are on equal footing in these observables.
The essential open problem is to define the Hilbert space of
states on which these observables act.  There is an important interplay
between states and time.

I would like to think J. York for helpful discussions.
This work was supported in part by National Science Foundation grant
PHY-9413207.


\begin{thebibliography}{99}
\bibitem{meas1} The adjective ``measurable'' and the verb ``to measure''
refer to the physical act of determining a value and are used
to avoid confusion and redundancy provoked by the words ``observable''
and ``to observe.''  They are not meant to invoke a specific interpretation
of quantum mechanics and in particular need not imply collapse of the
wave function.  The relation between various uses of the word
observable are discussed in Ref. \cite{And1}.
\end{thebibliography}
\end{document}